\documentclass[twocolumn,pra,amsmath,amssymb]{revtex4-1}
\usepackage{graphicx,bm,epsfig,color,tikz}

\begin{document}

%\title{{\bf Four electron spin qubits: stabilizer \& gates}}
\title{{\bf Four electron spin qubits with exchange interaction}}

\author{H.W.L. Naus}
%\author{H.W.L. Naus} \email{rik.naus@tno.nl}
\affiliation{Quantum Technology, TNO, P.O. Box 155, 2600 AD Delft and
QuTech, Delft University of Technology, P.O. Box 5046, 2600 GA Delft, The Netherlands}

\begin{abstract}
Four electron spin qubits in quantum dots are studied by means of an exchange interaction Hamiltonian.
The time-independent Schr\"odinger equation is exactly  analytically solved for the symmetric case,
that is equal qubit frequencies and equal couplings. The spin system can be used as two data-qubits and
two ancillas, thereby realizing a $X$ and $Z$ stabilizing circuit exploited in error correction and
 surface codes.
The unitary evolution of corresponding stable states is calculated and consequences
 for error syndromes and fidelities are assessed.
It is also shown that applying the evolution operator for a prescribed duration
 and subsequent measurement of the ancillas 
lead to a `verifiable' two-qubit phase gate.
\end{abstract}

\date{\today}
\maketitle

\newcommand{\ketz}{|\,\mathbf{0}\rangle}
\newcommand{\keto}{|\,\mathbf{1}\rangle}
\newcommand{\braz}{\langle\mathbf{0}\,|}
\newcommand{\brao}{\langle\mathbf{1}\,|}

\newcommand{\ketnz}{|\,n, \mathbf{0}\rangle}
\newcommand{\ketno}{|\,n, \mathbf{1}\rangle}
\newcommand{\branz}{\langle n, \mathbf{0}\,|}
\newcommand{\brano}{\langle n, \mathbf{1}\,|}

\newcommand{\siga}{\boldsymbol{\sigma}}
\newcommand{\sigc}{\boldsymbol{\sigma}^\dagger}

\newcommand{\ketu}{|\,\uparrow \, \rangle}
\newcommand{\ketd}{|\downarrow \, \rangle}
\newcommand{\brau}{\langle \, \uparrow |}
\newcommand{\brad}{\langle \, \downarrow|}

\newcommand{\ketuu}{|\,\uparrow \, \uparrow \, \rangle}
\newcommand{\ketud}{|\,\uparrow \, \downarrow \, \rangle}
\newcommand{\ketdu}{|\downarrow \, \uparrow \, \rangle}
\newcommand{\ketdd}{|\downarrow \, \downarrow \, \rangle}

\newcommand{\brauu}{\langle \, \uparrow \, \uparrow |}
\newcommand{\braud}{\langle \, \uparrow \, \downarrow|}
\newcommand{\bradu}{\langle \, \downarrow \, \uparrow |}
\newcommand{\bradd}{\langle \, \downarrow \, \downarrow|}

%\tableofcontents
%
%\newpage

\section{Introduction}
In this study, we consider four electron spin qubits in quantum dots, which interact
via an exchange Hamiltonian. Such a few-qubit system can serve as a simple stabilizing circuit
exploited in error correction for fault-tolerant quantum computing. It also may be used as unit
cell in a scalable many spin qubit quantum processor.

The reviews \cite{Engel,Hanson}, the thesis \cite{Kawa} and references therein
%the internal note  \cite{Naus-spin}
introduce the basics of electron spins in quantum dots. The essential ingredient is the description of
the low-energy physics of electrons in quantum dots by the exchange Hamiltonian, as argued in the 
seminal papers \cite{Loss98,Loss00}. The Heisenberg interaction for two confined electrons takes
their Coulomb interaction and the Pauli exclusion principle effectively into account \cite{Ash}. For electrons in quantum dots
the possible tunneling process is included in \cite{Meun,Watson} by enlarging the 4D two-spin Hilbert space to
a 6D space. In \cite{Meun} the effective exchange interaction is rederived. Note that turning the tunneling
on and off results in a time-dependent exchange coupling \cite{Loss98,Loss00}.
Here we restrict ourselves to constant coupling; if it is taken to be equal for the interacting pairs of spin,
the problem can be analytically exactly solved in the resonant case. Slightly different  qubit frequencies 
may be taken into account perturbatively; we also present the explicit form of this perturbation. 

After solving the model Hamiltonian, we address the unitary evolution of the stable states used in the $X$ and $Z$ stabilizer
circuits as introduced in \cite{Fowler}. Two qubits are considered as data qubits, whereas the other two are ancillas
which are supposed to be measured.  It is completely analogous to the recent analysis \cite{Naus-stab} of a coupled
qubit-resonator system. Of course, the degrees of freedom and their dynamics are different.
Therefore, we refer to \cite{Naus-stab} for further discussion and motivation of the followed approach. 
Our calculations have also demonstrated the possible implementation of phase gates for the data-qubits.
They may be called `checkable' or `verifiable' gates because of concomitant ancilla measurements.

The outline of this paper is as follows. First, the exchange Hamiltonian of four interacting spins
in a magnetic field is presented and solved for equal couplings and frequencies. In section  \ref{sec:stab},
the general final state of the stabilizing circuit is rewritten in terms of singlet-triplet states.
Subsequent unitary evolution and consequences for probabilities and fidelities are calculated in section \ref{sec:UE}.
The next section \ref{sec:Gate} demonstrates the concept of verifiable gates. The
final section presents conclusions. % Our notation is adopted from \cite{Naus-spin}.

\section{Four interacting spins}\label{sec:four}
In order to analyze a simple stabilizer \cite{Fowler}, we consider a spin system
consisting out of two data qubits $Q_1,Q_2$ and and two ancillas $Q_a,Q_b$.
They are arranged as shown in the figure and only nearest neighbour interaction,
indicated by the dashed red lines, is included.

\begin{center}
\begin{tikzpicture}

\draw [blue] (0,0) circle (0.5);
\node at (0,0) {$Q_1$};

\draw [blue] (4,0) circle (0.5);
\node at (4,0) {$Q_a$};

\draw [blue] (0,-4) circle (0.5);
\node at (0,-4) {$Q_b$};

\draw [blue] (4,-4) circle (0.5);
\node at (4,-4) {$Q_2$};

\draw [dashed,red,thick] (0,-.5) -- (0,-3.5); 
%\node at (-0.2,-2) {3};
\draw [dashed,red,thick] (4,-.5) -- (4,-3.5); 
%\node at (4.2,-2) {2};
\draw [dashed,red,thick] (0.5,0.) -- (3.5,0.); 
%\node at (2.0,0.3) {1};
\draw [dashed,red,thick] (0.5,-4.0) -- (3.5,-4.); 
%\node at (2.0,-4.3) {4};
\end{tikzpicture}
\end{center}

The model Hamiltonian incorporating the exchange interaction is taken as
\begin{eqnarray}
H &=& H_0 + H_{\text{int}} \nonumber \\
&=& \omega (S_z^{[1]} + S_z^{[2]})
+ \tilde \omega (S_z^{[a]} + S_z^{[b]}) \nonumber \\
 &+&  \mathcal{J} ( \vec S^{[1]} \cdot  \vec S^{[a]}
+ \vec S^{[1]} \cdot  \vec S^{[b]} + \vec S^{[2]} \cdot  \vec S^{[a]} + \vec S^{[2]} \cdot  \vec S^{[b]})\nonumber \\
 &=& \omega (S_z^{[1]} + S_z^{[2]})
+ \tilde \omega (S_z^{[a]} + S_z^{[b]}) \nonumber \\
  &+& \mathcal{J} (\vec S^{[1]}  +  \vec S^{[2]})
 \cdot  (\vec S^{[a]} + \vec S^{[b]}). 
\end{eqnarray}
The `free' Hamiltonian represents the energy of the electron spins in a 
homogeneous magnetic field in the $z$-direction. The electron spin operators are related to
the Pauli matrices
\begin{equation}
S_k = \tfrac{1}{2} \hbar \sigma_k, \quad k=x,y,z.
\end{equation}
The constant $\mathcal J$ is 
is commonly known as the exchange interaction strength.
After introducing the total four-spin operator $\vec J$ as
\begin{equation}
\vec J = \vec S^{[1]} + \vec S^{[2]} + \vec S^{[a]} + \vec S^{[b]} ,
\end{equation}
the interaction Hamiltonian can be rewritten as
\begin{equation}
H_{\text{int}} =  \tfrac{1}{2} \mathcal{J}\left(\vec J^2 - \left(\vec S^{[1]} + \vec S^{[2]}\right)^2
 - \left(\vec S^{[a]} + \vec S^{[b]}\right)^2\right). 
\end{equation}
It suggests the following construction of basis states. First, the data qubits are combined
as singlet and triplet states; the same is done for the ancillas. Thus we have defined the
respective two-qubit bases
\begin{eqnarray}
&&|S^{[12]}\rangle, |T_+^{[12]}\rangle, |T_0^{[12]}\rangle, |T_-^{[12]}\rangle \quad \text{and} \nonumber\\
&&|S^{[ab]}\rangle, |T_+^{[ab]}\rangle, |T_0^{[ab]}\rangle, |T_-^{[ab]}\rangle. 
\end{eqnarray}
Four-qubit basis states are now constructed by combining the two singlets, one singlet
with a triplet yielding spin 1 `particles' and the two triplets yielding
a spin 0, a spin 1 and a spin 2 `particle'. We explicitly obtain for the singlets
\begin{equation}
|e_1\rangle = |S^{[12]}\rangle \otimes |S^{[ab]}\rangle \qquad (j,m_j)=(0,0)
\end{equation}
and for the singlet triplet combinations
\begin{eqnarray}
|e_2\rangle &=& |S^{[12]}\rangle  \otimes |T_+^{[ab]}\rangle \qquad (j,m_j)=(1,1), \nonumber\\
|e_3\rangle &=& |S^{[12]}\rangle  \otimes |T_0^{[ab]}\rangle \qquad (j,m_j)=(1,0), \nonumber\\
|e_4\rangle &=& |S^{[12]}\rangle  \otimes |T_-^{[ab]}\rangle \qquad(j,m_j)=(1,-1), \nonumber\\
|e_5\rangle &=& |T_+^{[12]}\rangle  \otimes |S^{[ab]}\rangle \qquad (j,m_j)=(1,1), \nonumber\\
|e_6\rangle &=& |T_0^{[12]}\rangle  \otimes |S^{[ab]}\rangle \qquad (j,m_j)=(1,0), \nonumber\\
|e_7\rangle &=& |T_-^{[12]}\rangle  \otimes |S^{[ab]}\rangle \qquad (j,m_j)=(1,-1). 
\end{eqnarray}
We have indicated the quantum numbers of the total spin of the states.
Combining the triplets yields \cite{Cohen2} the scalar
\begin{widetext}
\begin{equation}
|e_8\rangle = \tfrac{1}{3} \sqrt{3}
\left\{|T_+^{[12]}\rangle \otimes |T_-^{[ab]}\rangle
-|T_0^{[12]}\rangle \otimes |T_0^{[ab]}\rangle 
+ |T_-^{[12]}\rangle \otimes |T_+^{[ab]}\rangle\right\}
\qquad (j,m_j)=(0,0)
\end{equation}
%\begin{eqnarray}
%|e_8\rangle &=& \tfrac{1}{3} \sqrt{3}
%\left\{|T_+^{[12]}\rangle \otimes |T_-^{[ab]}\rangle
%-|T_0^{[12]}\rangle \otimes |T_0^{[ab]}\rangle \right.\nonumber\\
%&&\left. + |T_-^{[12]}\rangle \otimes |T_+^{[ab]}\rangle\right\}
%\qquad (j,m_j)=(0,0)
%\end{eqnarray}
and the spin 1 states
%\begin{widetext}
\begin{eqnarray}
|e_9\rangle &=& \tfrac{1}{2}\sqrt{2}\left\{
|T_+^{[12]}\rangle  \otimes |T_0^{[ab]}\rangle -|T_0^{[12]}\rangle  \otimes |T_+^{[ab]}\rangle\right\}
\quad (j,m_j)=(1,1), \nonumber\\
|e_{10}\rangle &=&  \tfrac{1}{2}\sqrt{2}\left\{
|T_+^{[12]}\rangle  \otimes |T_-^{[ab]}\rangle -|T_-^{[12]}\rangle  \otimes |T_+^{[ab]}\rangle\right\}
\quad (j,m_j)=(1,0), \nonumber\\
|e_{11}\rangle &=&  \tfrac{1}{2}\sqrt{2}\left\{
|T_0^{[12]}\rangle  \otimes |T_-^{[ab]}\rangle -|T_-^{[12]}\rangle  \otimes |T_0^{[ab]}\rangle\right\}
\quad (j,m_j)=(1,-1).
\end{eqnarray}
Finally, we get the resulting spin 2 states
\begin{eqnarray}
|e_{12}\rangle &=& |T_+^{[12]}\rangle \otimes |T_+^{[ab]}\rangle \qquad (j,m_j)=(2,2),\nonumber \\
|e_{13}\rangle &=& \tfrac{1}{2}\sqrt{2}\left\{
|T_+^{[12]}\rangle  \otimes |T_0^{[ab]}\rangle +|T_0^{[12]}\rangle  \otimes |T_+^{[ab]}\rangle\right\}
\qquad (j,m_j)=(2,1), \nonumber\\
|e_{14}\rangle &=& \tfrac{1}{6} \sqrt{6}
\left\{|T_+^{[12]}\rangle \otimes |T_-^{[ab]}\rangle
+2|T_0^{[12]}\rangle \otimes |T_0^{[ab]}\rangle
+ |T_-^{[12]}\rangle \otimes |T_+^{[ab]}\rangle\right\}
\quad (j,m_j)=(2,0), \nonumber\\
|e_{15}\rangle &=& \tfrac{1}{2}\sqrt{2}\left\{
|T_0^{[12]}\rangle  \otimes |T_-^{[ab]}\rangle +|T_-^{[12]}\rangle  \otimes |T_0^{[ab]}\rangle\right\}
\qquad (j,m_j)=(2,-1), \nonumber\\
|e_{16}\rangle &=& |T_-^{[12]}\rangle \otimes |T_-^{[ab]}\rangle \qquad (j,m_j)=(2,-2).
\end{eqnarray}
\end{widetext}

If all qubits have the same frequency $\omega$ then the Hamiltonian $H_0$  reduces to
\begin{equation}
H_0= \omega J_z,
\end{equation}
{\em i.e.}, proportional to the $z$-component of the total spin. Recall that we have presupposed
equal exchange couplings from the onset.
In this highly symmetric case, the complete Hamiltonian is diagonal in the defined basis $|e_k\rangle, k=1,2\dots 16$.
This is due to the fact the basis states are eigenstates of $\vec J^2$ and $J_z$ but also
of $ (\vec S^{[1]}+\vec S^{[2]})^2$ and
$(\vec S^{[a]}+\vec S^{[b]})^2$; these operators commute with $H$ for $\omega=\tilde \omega$.
In other words, we have solved the time-independent Schr\"odinger equation for this case.
The diagonal elements, these are the resonant eigenenergies, are found as $H_{kk}=h_k$, with
\begin{eqnarray}
\label{eq:ener}
h_{1} &=& 0, \quad h_{2} = \hbar \omega, \quad h_{3} = 0, \quad h_{4} = -\hbar\omega,\nonumber\\ 
h_{5} &=& \hbar\omega, \quad h_{6} = 0, \quad h_{7} = -\hbar\omega, \quad h_{8} = -2 \hbar^2\mathcal J, \nonumber\\ 
h_{9} &=&\hbar\omega - \hbar^2\mathcal J, \quad h_{10} =- \hbar^2\mathcal J, \quad h_{11} = -\hbar\omega-  \hbar^2\mathcal J,\nonumber\\ 
\quad h_{12} &=& 2\hbar\omega +  \hbar^2\mathcal J,
h_{13} =\hbar\omega + \hbar^2\mathcal J, \quad h_{14} = \hbar^2\mathcal J,\nonumber \\ h_{15} &=& -\hbar\omega+  \hbar^2\mathcal J, \quad h_{16} = - 2\hbar\omega + \hbar^2 \mathcal J.
\end{eqnarray}
This result can be verified by explicit calculation.

For nonequal frequencies $\omega, \tilde \omega$ the free Hamiltonian gets an additional term
\begin{equation}
H_0= \omega J_z -\delta\omega(S_z^{[a]}+S_z^{[b]}),
\end{equation}
with frequency difference $\delta \omega = \omega-\tilde \omega$.
Some diagonal matrix elements get modified, $h_k \rightarrow \tilde h_k$:
\begin{eqnarray}
\tilde h_{2}&=& \hbar(\omega-\delta \omega), \quad \tilde h_{4} = -\hbar(\omega - \delta \omega),\nonumber \\ 
\tilde h_{9} &=&\hbar(\omega-\tfrac{1}{2}\delta\omega) - \hbar^2\mathcal J,  \quad \tilde h_{11} = -\hbar(\omega-\tfrac{1}{2}\delta\omega)-  \hbar^2\mathcal J,\nonumber \\
\tilde h_{12} &=& 2\hbar\omega-\hbar\delta\omega +  \hbar^2\mathcal J, 
\tilde h_{13} =\hbar(\omega-\tfrac{1}{2}\delta\omega) + \hbar^2\mathcal J,  \\ \tilde h_{15} &=& -(\hbar\omega-\tfrac{1}{2}\delta\omega)+  \hbar^2\mathcal J,
\quad \tilde h_{16} = - 2\hbar\omega + \hbar\delta\omega +  \hbar^2\mathcal J. \nonumber 
\end{eqnarray}
We furthermore obtain the following non-diagonal terms
\begin{eqnarray}
H_{8-10} &=& \tfrac{1}{3}\sqrt{6} \hbar \delta \omega, \quad H_{9-13}= \tfrac{1}{2} \hbar \delta \omega, \nonumber \\
H_{10-14} &=& \tfrac{1}{3}\sqrt{3} \hbar \delta \omega, \quad H_{11-15} = \tfrac{1}{2} \hbar \delta \omega 
\end{eqnarray}
and the identical transposed matrix elements.
The consequence is
mixing between the three states $|e_8\rangle, |e_{10}\rangle, |e_{14}\rangle$, between the two states
$|e_9\rangle, |e_{13}\rangle$ and between $|e_{11}\rangle, |e_{15}\rangle$.

\section{Stabilizing circuit}\label{sec:stab}
In analogy to \cite{Naus-stab}, we analyze the stabilizing four-qubit circuit presented in \cite{Fowler},
which is used in error
correction surface codes. In particular, we study the consequences of unitary evolution
 governed by the Hamiltonian describing the interacting spins.
To this end, we first rewrite the final state of the stabilizer circuit after putting in a general two-data qubit state with
the ancillas in the ground state as given in \cite{Fowler,Naus-stab} as
\begin{eqnarray}
|\psi\rangle
&=&  A_+ |\Phi^+ \rangle \otimes \ketdd 
+ B_+ |\Psi^+ \rangle  \otimes \ketdu \nonumber \\
&+&  A_- |\Phi^- \rangle  \otimes\ketud
+ B_- |\Psi^- \rangle   \otimes\ketuu \nonumber \\
&=& \tfrac{1}{2}\sqrt{2} A_+ (|T_+^{[12]} \rangle + |T_-^{[12]}\rangle) \otimes \ketdd \nonumber \\
&+& B_+ |T_0^{[12]} \rangle  \otimes \ketdu - B_- |S^{[12]} \rangle   \otimes\ketuu
 \nonumber \\
&+&  \tfrac{1}{2}\sqrt{2} A_- (|T_-^{[12]} \rangle - |T_+^{[12]}\rangle)  \otimes\ketud.
\end{eqnarray}
Here we have introduced the data-qubit Bell states
\begin{eqnarray}
|\Phi^\pm \rangle &=& \tfrac{1}{2}(\ketdd \pm \ketuu), \nonumber \\
|\Psi^\pm \rangle  &=&\tfrac{1}{2}(\ketdu \pm \ketud) .
\end{eqnarray}
Expressing the ancilla states in singlet-triplet states yields
\begin{eqnarray}
|\psi\rangle &=& \tfrac{1}{2}\sqrt{2}A_+(|T_+^{[12]} \rangle + |T_-^{[12]}\rangle) \otimes |T_-^{[ab]}\rangle \nonumber\\
&+&\tfrac{1}{2}\sqrt{2} B_+ |T_0^{[12]} \rangle  \otimes (|T_0^{[ab]}\rangle-|S^{[ab]}\rangle)   \nonumber \\
&+&  \tfrac{1}{2}A_- (|T_-^{[12]} \rangle - |T_+^{[12]}\rangle)  \otimes (|S^{[ab]}\rangle +|T_0^{[ab]}\rangle) \nonumber \\
&-& B_- |S^{[12]} \rangle   \otimes  |T_+^{[ab]}\rangle.
\label{eq:genstate}
\end{eqnarray}
Measuring the ancillas projects this state onto the product of a data-qubit state and the ancilla state
corresponding to the obtained result. Re-inserting such a state into the circuit, including re-initialization
of the ancillas in the ground state, yields the same state. It means that the measurement result
is again found with probability one. The state is therefore called stable. Unitary evolution, however,
may reduce the probability to obtain the desired result. This would be interpreted as an error 
detection \cite{Fowler, Naus-stab}.
Apart from the changed, that is non-zero and not one probabilities, the fidelity of the state after
some elapsed time is a measure for the unwanted effects of evolution. 

\section{Unitary evolution and its consequences}\label{sec:UE}
Here we consider the unitary evolution for the case of zero detuning, {\it i.e.}, $\omega = \tilde \omega$.
As shown above, it implies that the introduced basis states $|e_k \rangle$ are the eigenstates of the Hamiltonian
with eigenenergies $E_k=h_k$.
The evolution operator  therefore follows as
\begin{equation}
U(t) = \sum_{k=1}^{16} \exp{(-iE_k t)} |e_k\rangle \langle e_k|,
\label{eq:evo}
\end{equation}
where we have adopted $\hbar =1$.
The ancilla measurement operators  are given by
\begin{eqnarray}
P_{\downarrow \downarrow} &=& \mathcal I^{12} \otimes \ketdd \bradd 
= \mathcal I^{12} \otimes |T_-^{[ab]}\rangle \langle T_-^{[ab]} |, \nonumber \\
P_{\downarrow \uparrow} &=& \mathcal I^{12} \otimes \ketdu \bradu \nonumber \\
&=& \tfrac{1}{2} \mathcal I^{12} \otimes (|T_0^{[ab]}\rangle -|S^{[ab]}\rangle)
(\langle T_0^{[ab]} |-\langle S^{[ab]}|), \nonumber \\
P_{\uparrow \downarrow} &=& \mathcal I^{12} \otimes \ketud \braud \nonumber \\ 
&=& \tfrac{1}{2} \mathcal I^{12} \otimes (|T_0^{[ab]}\rangle +|S^{[ab]}\rangle)
(\langle T_0^{[ab]} |+\langle S^{[ab]}|), \nonumber \\
P_{\uparrow \uparrow} &=& \mathcal I^{12} \otimes \ketuu \brauu 
= \mathcal I^{12} \otimes |T_+^{[ab]}\rangle \langle T_+^{[ab]} |.
\label{eq:meop}
\end{eqnarray}
In the following we calculate the consequences of evolution for the four different `initial' states.
We denote them as in \cite{Fowler, Naus-stab}.
The shown numerical results are calculated with the parameter values: $\omega=18.5$ GHz
and $\mathcal J=80$  MHz, $\mathcal J=800$  MHz.

\subsection{Bell state $|\Phi^+\rangle$}
\begin{figure}[htb] 
\centering
\includegraphics[width=7.9cm]{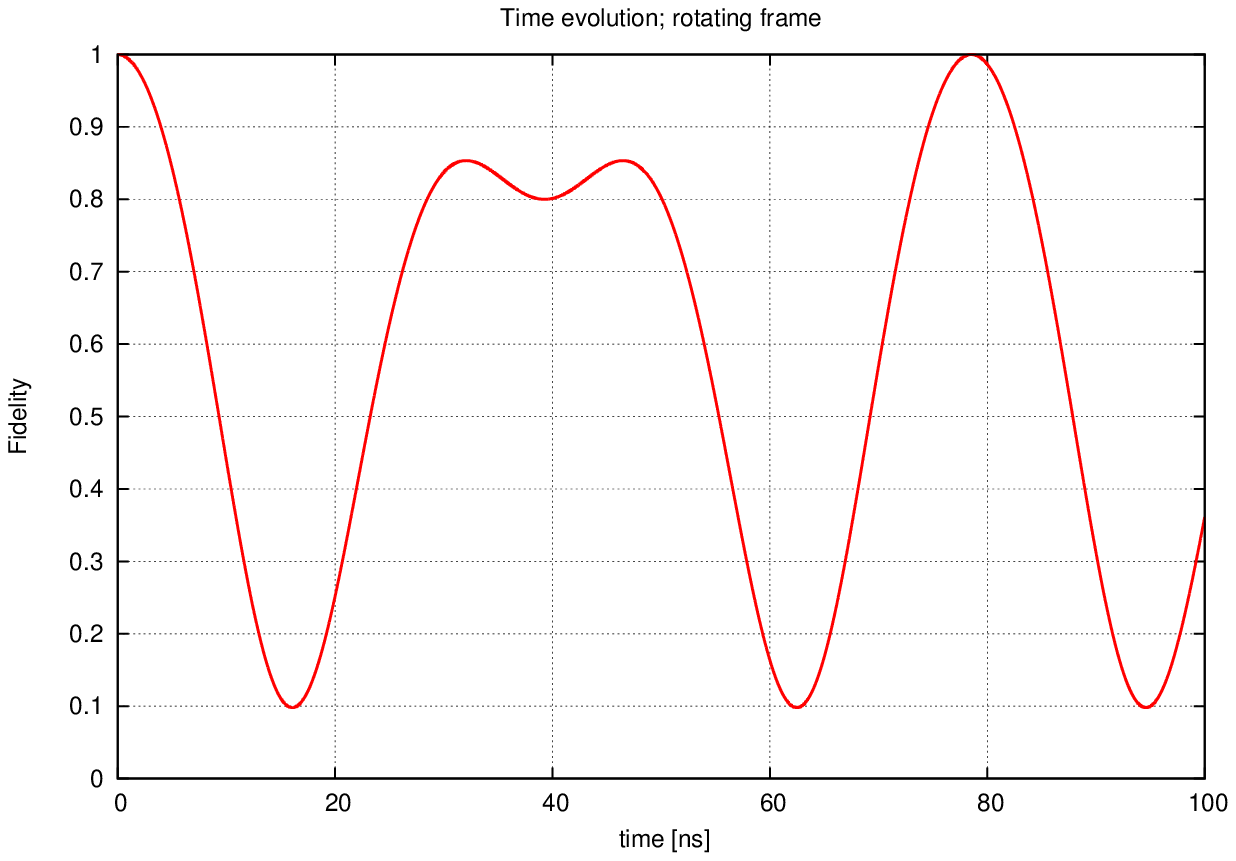}
\includegraphics[width=7.9cm]{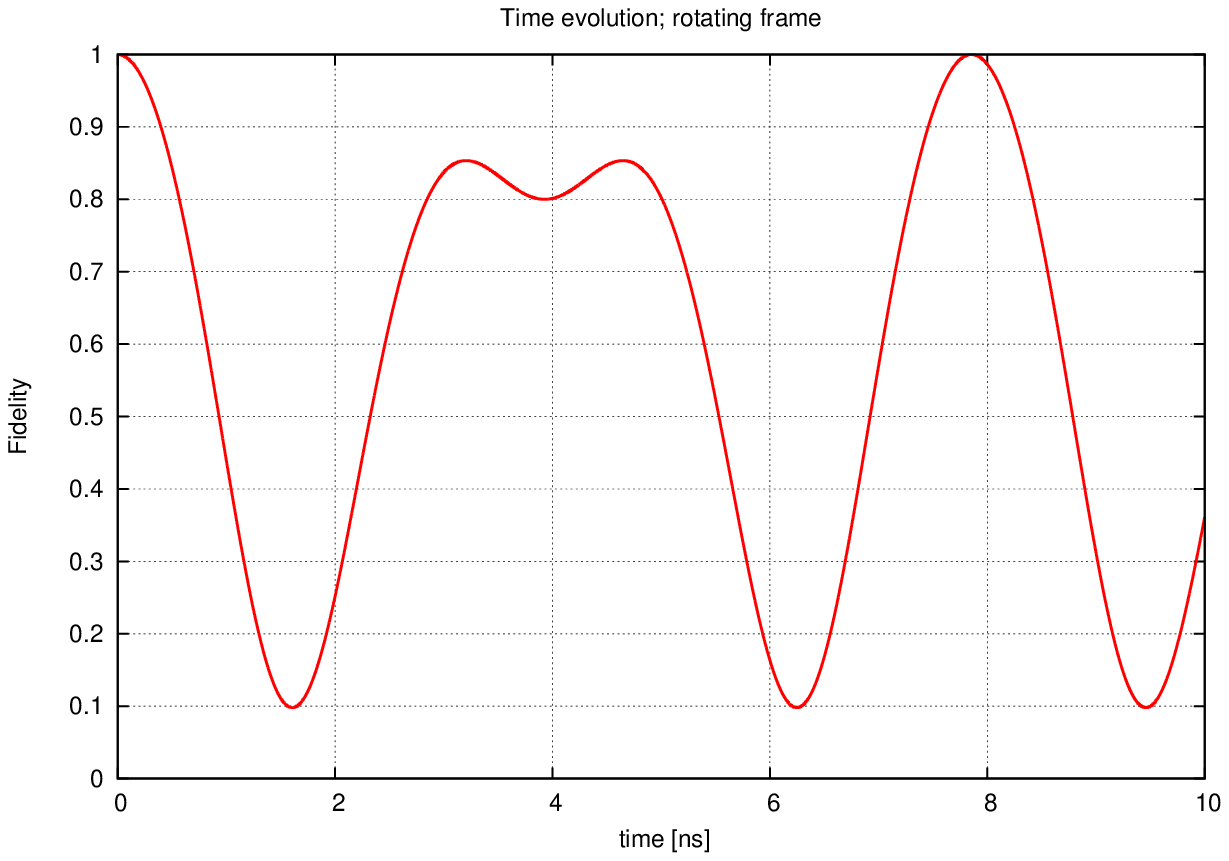}
\caption{\label{fig:fid1} Fidelities: top $\mathcal J=$ 80 MHz; low $\mathcal J=$ 800 MHz; note the timescales.}
\end{figure}

The corresponding four qubit  state $|\phi^+\rangle$
follows from (\ref{eq:genstate}) by taking $A_+=1, A_-=B_+=B_-=0$. 
We expand this state in terms of the basis states and find
\begin{equation}
|\phi^+\rangle = \tfrac{1}{6}\sqrt{6} |e_8\rangle +
\tfrac{1}{2} |e_{10}\rangle + \tfrac{1}{6} \sqrt{3} |e_{14}\rangle + \tfrac{1}{2} \sqrt{2} |e_{16}\rangle.
\end{equation}
Its time evolution is easily obtained as
\begin{eqnarray}
|\phi^+(t)\rangle &=& \tfrac{1}{6}\sqrt{6} e^{-iE_8t} |e_8\rangle +
 \tfrac{1}{2}  e^{-iE_{10}t}|e_{10}\rangle \nonumber \\ &+& \tfrac{1}{6} \sqrt{3}  e^{-iE_{14}t}|e_{14}\rangle +
\tfrac{1}{2} \sqrt{2}  e^{-iE_{16}t}|e_{16}\rangle.
\end{eqnarray}
By means of (\ref{eq:meop}) we then get
\begin{equation}
P_{\downarrow \downarrow}|\phi^+(t)\rangle =
\tfrac{1}{2} \sqrt{2} \left\{\alpha(t)
 |T_+^{[12]} \rangle +
e^{-iE_{16}t}|T_-^{[12]}\rangle\right\} \otimes |T_-^{[ab]}\rangle,
\end{equation}
where we have defined
\begin{equation}
\alpha(t)= \tfrac{1}{3} e^{-iE_8t}  +
 \tfrac{1}{2}  e^{-iE_{10}t} + \tfrac{1}{6} e^{-iE_{14}t}.
\end{equation}
The probability of indeed measuring $\downarrow \downarrow$ therefore reads
\begin{equation}
p_{\downarrow \downarrow}= 
\tfrac{1}{2}\left(1 + \alpha^*(t) \alpha(t)\right).
\end{equation}
After factoring out the ancillas we obtain as normalized final two-qubit state
\begin{equation}
|\Phi(t)\rangle = \frac{1}
{\sqrt{2p_{\downarrow \downarrow}}} \left\{\alpha(t)
 |T_+^{[12]} \rangle + e^{-iE_{16}t}|T_-^{[12]}\rangle\right\}.
\end{equation}
This leads to the fidelity
\begin{eqnarray}
&&F= |\langle \Phi^+|\Phi(t)\rangle |  \\ =
&&\frac{1}{2} \left[
1+ \frac{1}{1+\alpha^*(t)\alpha(t)}\left(\alpha(t)e^{iE_{16}t}+\alpha^*(t)e^{-iE_{16}t}\right)
\right]^{1/2}.\label{eq:fid1} \nonumber
\end{eqnarray}
It turns out that is a very fast oscillating quantity, governed by the frequency $2\omega$.
This less interesting `free' evolution can be eliminated exploiting the rotating frame, cf. \cite{Naus-stab}. 
The fidelity in this rotating frame is obtained by replacing $E_{16}$ by $\mathcal J$ in (\ref{eq:fid1}) and
is depicted in figure (\ref{fig:fid1}) for two values of the exchange constant $\mathcal J$. 
Changing the vales of $\mathcal J$ merely changes the timescale: stronger coupling causes a faster
decrease in fidelity.

\begin{figure}[htb] 
\centering
\includegraphics[width=7.9cm]{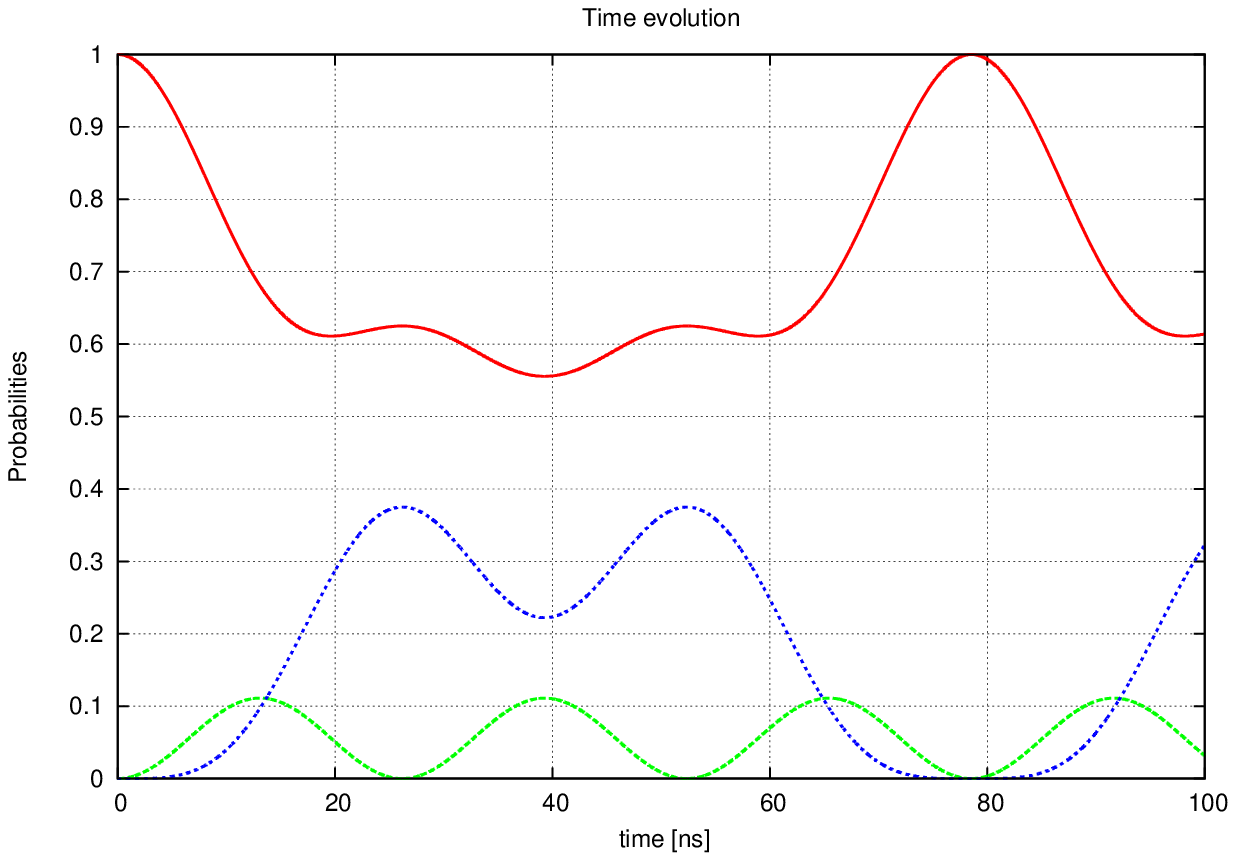}
\includegraphics[width=7.9cm]{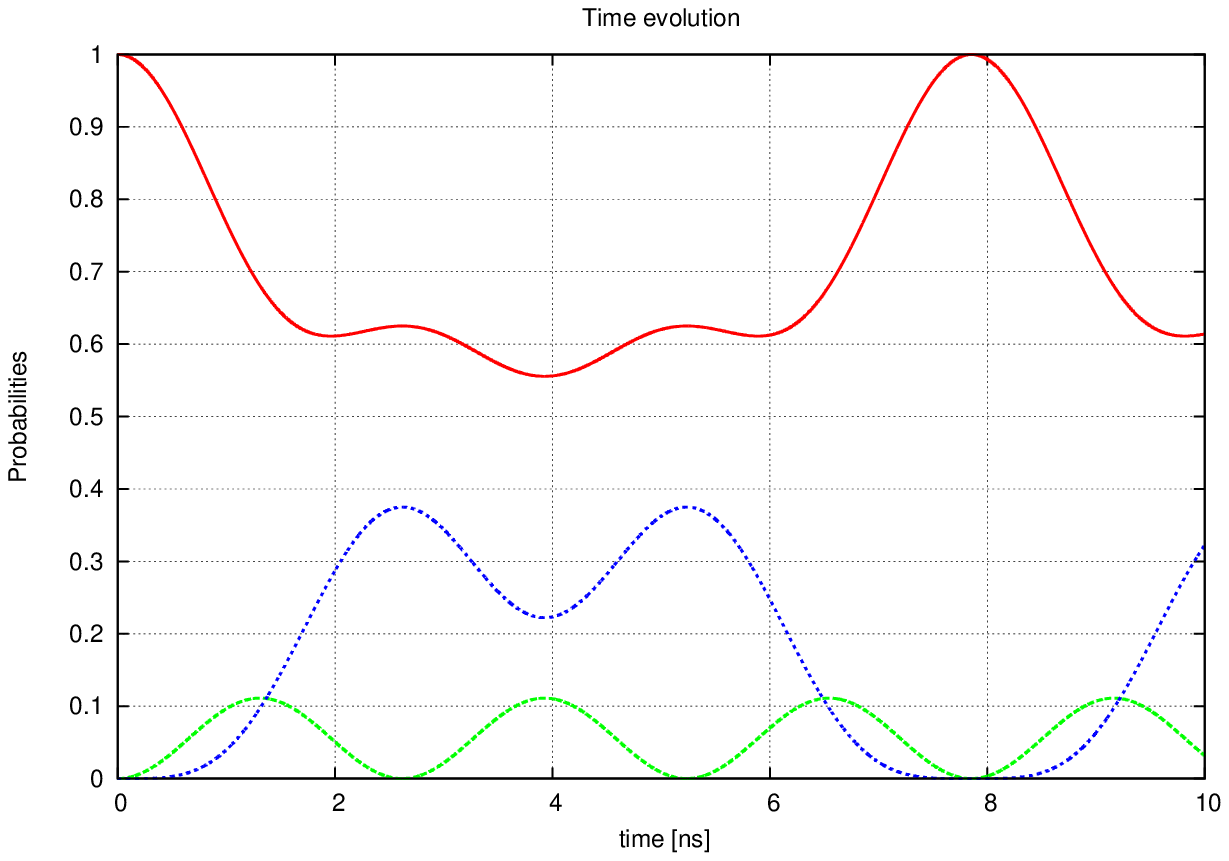}
\caption{\label{fig:evo1} Probabilities:
$p_{\downarrow \downarrow}$ red,
$p_{\uparrow \downarrow}= p_{\downarrow \uparrow}$ green,
$p_{\uparrow \uparrow}$ blue;
top $\mathcal J=$ 80 MHz; low $\mathcal J=$ 800 MHz; note the timescales.}
\end{figure}

Analogous computations yield the probabilities to obtain the other possible measurement results:
\begin{equation}
p_{\downarrow \uparrow}= p_{\uparrow \downarrow}=
\tfrac{1}{18}\left[1-\cos{\left\{(E_{14}-E_8)t\right\}}\right]
\end{equation}
and
\begin{equation}
p_{\uparrow \uparrow}= \beta^*(t)\beta(t),
\end{equation}
 with 
\begin{equation}
\beta(t) = \tfrac{1}{2}\sqrt{2}\left(\tfrac{1}{3} e^{-iE_8t}  
- \tfrac{1}{2} e^{-iE_{10}t} + \tfrac{1}{6}  e^{-iE_{14}t}\right).
\end{equation}
Note that the sum of the probabilities indeed is equal to one for all $t$.  
They are shown in figure (\ref{fig:evo1}).
Again we note that altering the value of $\mathcal J$ changes the timescale; in this case for the varying  probabilities.

\subsection{Bell state $|\Psi^+\rangle$}
We repeat these computations for the other Bell states as well. First, we consider $|\Psi^+\rangle$ which 
corresponds to $B_+=1, A_+=A_-=B_-=0$ in (\ref{eq:genstate}).
Initial and time-evolved state follow respectively as
\begin{equation}
|\psi^+\rangle = -\tfrac{1}{2}\sqrt{2} |e_6\rangle 
- \tfrac{1}{6} \sqrt{6} |e_{8}\rangle + \tfrac{1}{3}\sqrt{3} |e_{14}\rangle
\end{equation}
and
\begin{eqnarray}
|\psi^+(t)\rangle &=& -\tfrac{1}{2}\sqrt{2} e^{-iE_{6}t} |e_6\rangle 
- \tfrac{1}{6} \sqrt{6}  e^{-iE_{8}t}|e_{8}\rangle \nonumber\\ &+& \tfrac{1}{3}\sqrt{3}  e^{-iE_{14}t}|e_{14}\rangle.
\end{eqnarray}
Using the measurement operator corresponding to the `correct result' yields
\begin{equation}
P_{\downarrow \uparrow}|\psi^+(t)\rangle = \gamma(t)
 |T_0^{[12]} \rangle \otimes (|T_0^{[ab]}\rangle - |S^{[ab]}\rangle),
\end{equation}
a product state, with time dependence 
\begin{equation}
 \gamma(t)= \tfrac{1}{2} e^{-iE_6t}  +
 \tfrac{1}{6} e^{-iE_{8}t} + \tfrac{1}{3}   e^{-iE_{14}t}.
\end{equation}
This leads to the probability
\begin{equation}
p_{\downarrow \uparrow}= 
\gamma^*(t) \gamma(t).
\end{equation}
The resulting two-qubit state
\begin{equation}
|\Psi(t)\rangle = \frac{\gamma(t)}{\sqrt{p_{\downarrow \uparrow}}} 
 |T_0^{12]} \rangle 
\end{equation}
has fidelity one with respect to the target state $|\Psi^+\rangle$.

\begin{figure}[htb] 
\centering
\includegraphics[width=7.9cm]{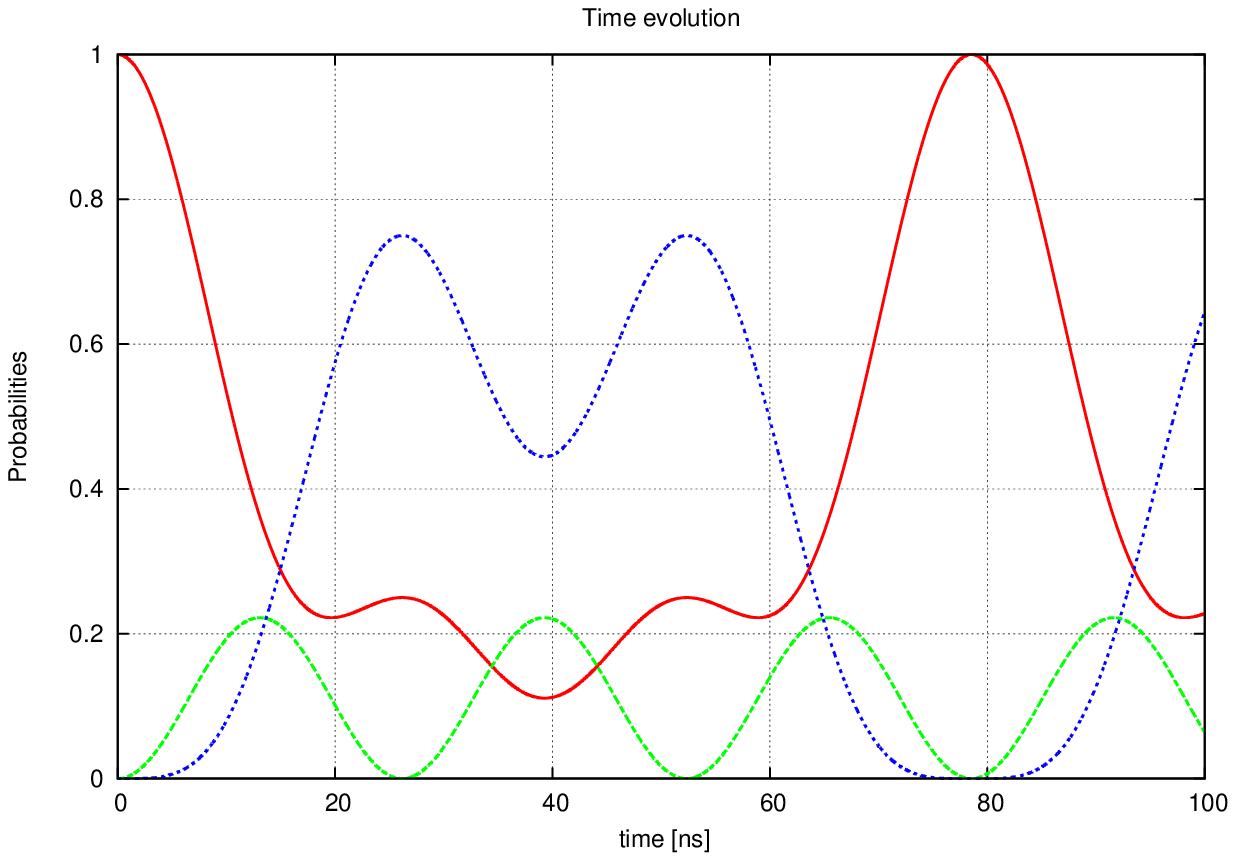}
\includegraphics[width=7.9cm]{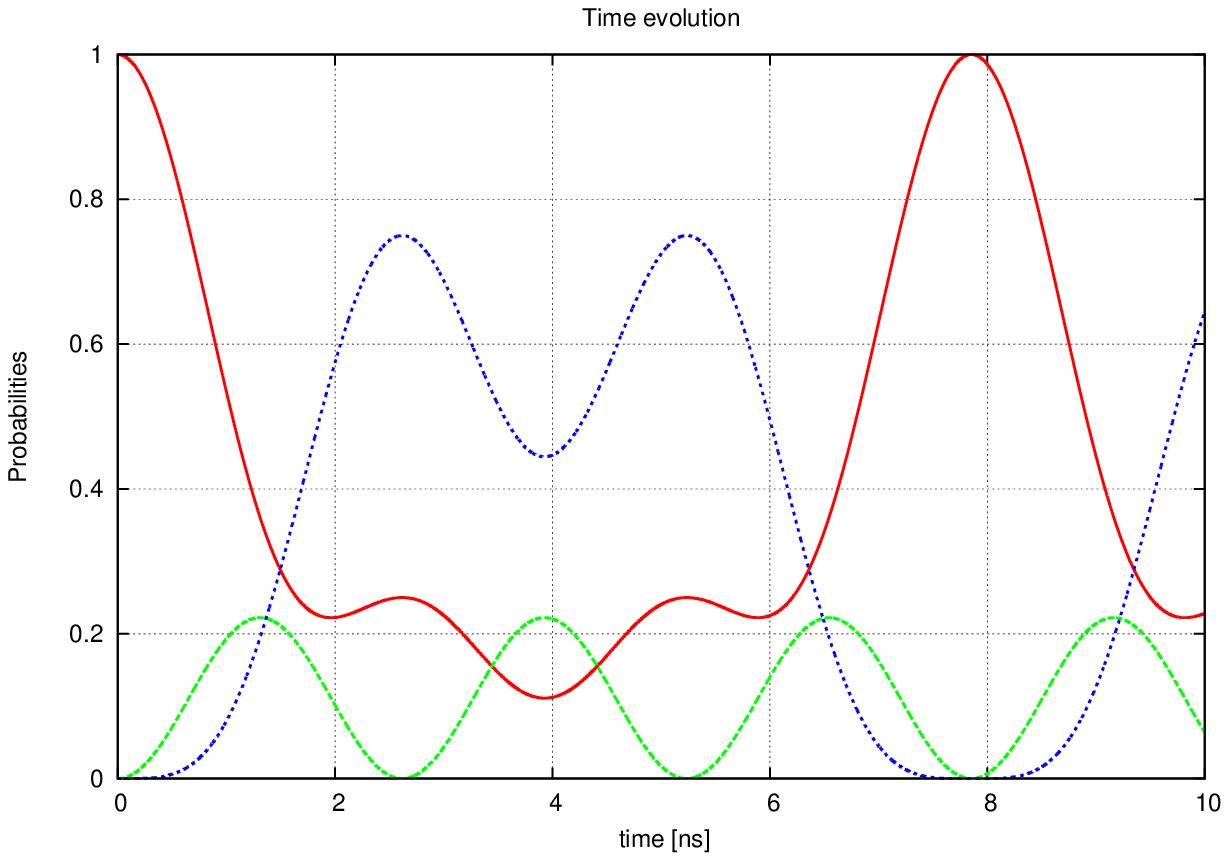}
\caption{\label{fig:evo2} Probabilities:
$p_{\downarrow \uparrow}$ red,
$p_{\uparrow \uparrow}= p_{\downarrow \downarrow}$ green,
$p_{\uparrow \downarrow}$ blue;
top $\mathcal J=$ 80 MHz; low $\mathcal J=$ 800 MHz; note the timescales.}
\end{figure}

The probabilities for obtaining the results which would be interpreted as error
detection follow as 
\begin{equation}
p_{\downarrow \downarrow}= p_{\uparrow \uparrow}=
\tfrac{1}{9}\left[1-\cos{\left\{(E_{14}-E_8)t\right\}}\right]
\end{equation}
and
\begin{equation}
p_{\uparrow \downarrow}= \mu^*(t)\mu(t),
\end{equation}
with
\begin{equation}
\mu(t) =-\tfrac{1}{2} e^{-iE_6t}  
+ \tfrac{1}{6} e^{-iE_{8}t} + \tfrac{1}{3}  e^{-iE_{14}t}.
\end{equation}
The various probabilities, indeed adding up to one, are shown in figure (\ref{fig:evo2}).
Once more, we see time-scaling with the coupling strength.

\subsection{Bell state $|\Phi^-\rangle$}
Next, we take $A_-=1, A_+=B_-=B_+=0$ in (\ref{eq:genstate}) corresponding to the two-qubit state
$|\Phi^-\rangle$.  
We immediately proceed to the state after time-evolution: 
\begin{eqnarray}
|\phi^-(t)\rangle &=& -\tfrac{1}{2} e^{-iE_{5}t} |e_5\rangle +
 \tfrac{1}{2}  e^{-iE_{7}t}|e_{7}\rangle - \tfrac{1}{4}\sqrt{2}  e^{-iE_{9}t}|e_{9}\rangle \nonumber \\
 &-& \tfrac{1}{4}\sqrt{2} e^{-iE_{11}t} |e_{11}\rangle 
 - \tfrac{1}{4}\sqrt{2} e^{-iE_{13}t} |e_{13}\rangle  \nonumber \\
 &+& \tfrac{1}{4}\sqrt{2} e^{-iE_{14}t} |e_{14}\rangle.
\end{eqnarray}
Operating with the appropriate measurement operator for the `correct result' gives
\begin{eqnarray}
P_{\uparrow \downarrow}|\phi^-(t)\rangle &=& -\tfrac{1}{8}\sqrt{2}\left\{
 \chi_+(t) |T_+^{[12]} \rangle - \chi_-(t) |T_-^{[12]} \rangle \right\} \nonumber\\
&\otimes& \left(|T_0^{[ab]}\rangle + |S^{[ab]}\rangle\right),
\end{eqnarray}
with the functions 
\begin{eqnarray}
\chi_+(t) &=& 2e^{-iE_5t}  + e^{-iE_{9}t} + e^{-iE_{13}t}, \nonumber \\
\chi_-(t) &=& 2e^{-iE_7t}  + e^{-iE_{11}t} + e^{-iE_{15}t}.
\end{eqnarray}
The probability to get the measurement result $\uparrow, \downarrow$ follows as
\begin{equation}
p_{\uparrow \downarrow}= \tfrac{1}{32}\left(\chi^*_+(t)\chi_+(t)
+ \chi^*_-(t)\chi_-(t)\right).
\end{equation}
After factoring out the ancilla states
we obtain for the corresponding fidelilty
\begin{equation}
F= \tfrac{1}{8}
\left[ \left(\chi_+(t)+\chi_-(t)\right) \left(\chi^*_+(t)+\chi^*_-(t)\right)/p_{\uparrow \downarrow}\right]^{1/2}.
\end{equation}
This fidelity also oscillates very fast. This is a result of the interference terms $\chi_+^* \chi_-$ and $\chi_-^* \chi^+$
which have a frequency of $2 \omega$. In the rotating frame of the qubits it vanishes.
The resulting fidelity in this frame is equal one for all times since in the rotating frame  
\begin{equation}
\chi_\pm(t) \rightarrow \tilde \chi_\pm(t) = e^{\pm i\omega t} \chi_\pm (t)
\end{equation}
and $\tilde \chi_+(t)= \tilde \chi_-(t)$. 
As a consequence, we obtain
\begin{equation}
F= \tfrac{1}{8}
\left[ 64p_{\uparrow \downarrow} /p_{\uparrow \downarrow}\right]^{1/2}= 1,
\end{equation}
which may be somewhat surprising.

\begin{figure}[htb] 
\centering
\includegraphics[width=7.9cm]{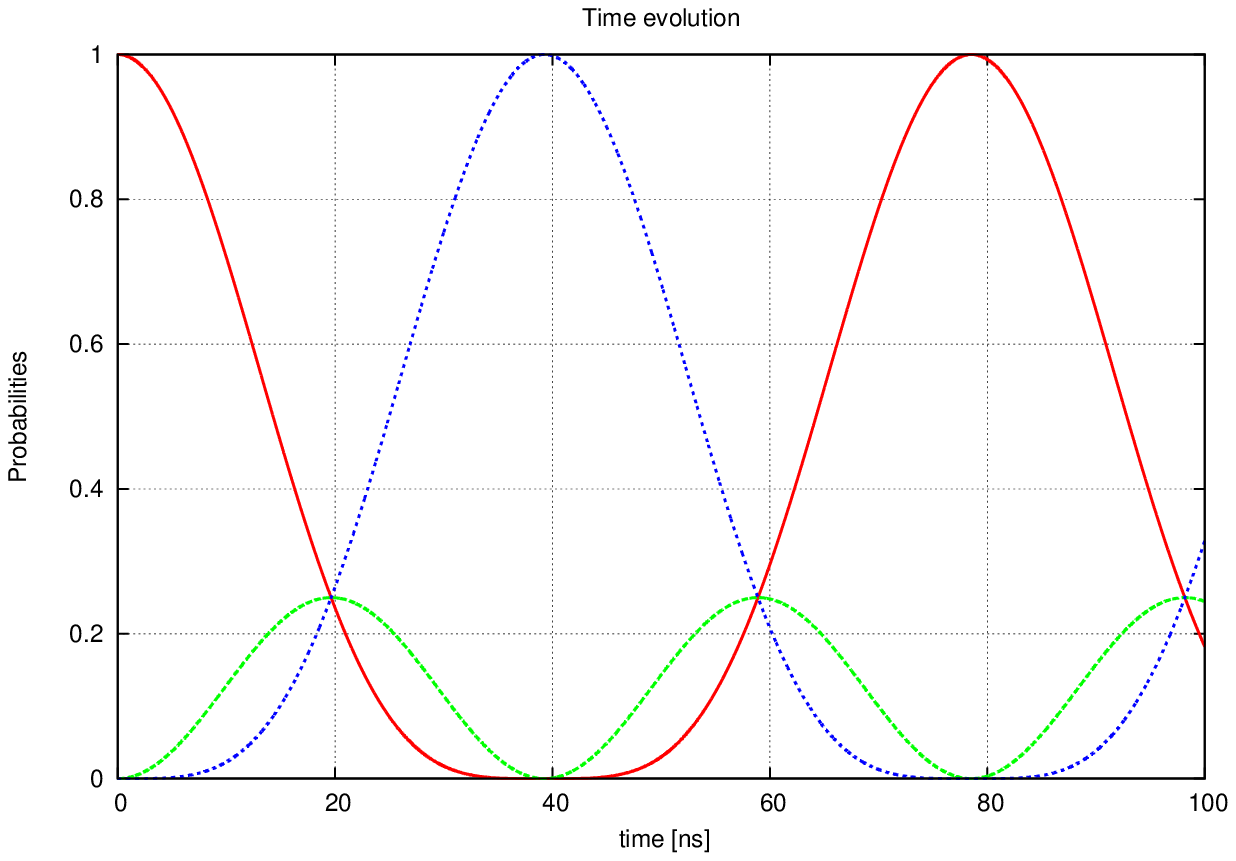}
\includegraphics[width=7.9cm]{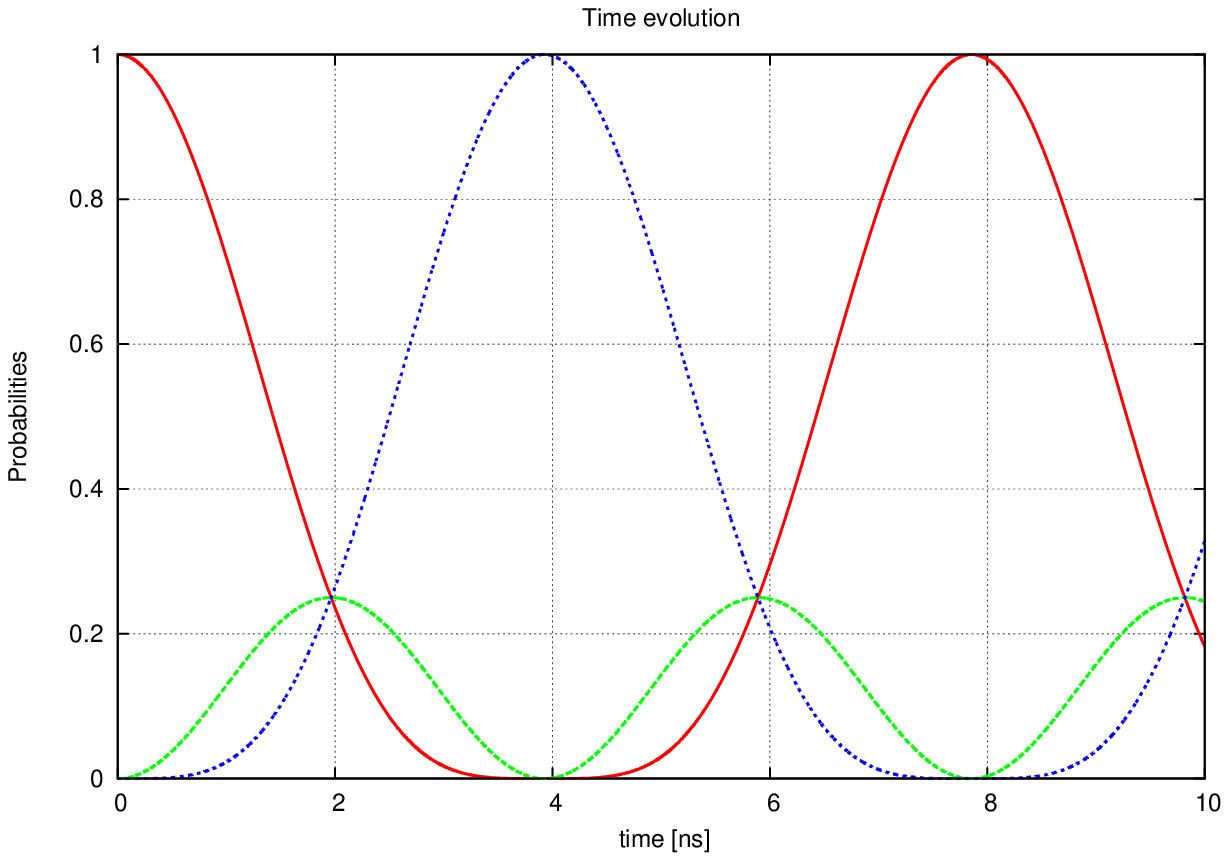}
\caption{\label{fig:evo3} Probabilities:
$p_{\uparrow \downarrow}$ red,
$p_{\uparrow \uparrow}= p_{\downarrow \downarrow}$ green,
$p_{\downarrow \uparrow}$ blue;
top $\mathcal J=$ 80 MHz; low $\mathcal J=$ 800 MHz; note the timescales.}
\end{figure}

The other, `error detection' probabilities can be calculated analogously. We get the results
\begin{eqnarray}
p_{\downarrow \downarrow} &=& \tfrac{1}{8}
\left[1-\cos{\left\{(E_{15}-E_{11})t\right\}}\right], \nonumber \\
p_{\uparrow \uparrow} &=& \tfrac{1}{8}
\left[1-\cos{\left\{(E_{13}-E_{9})t\right\}}\right], \nonumber \\
p_{\downarrow \uparrow} &=& \tfrac{1}{32}\left(\nu^*_+(t)\nu_+(t)
+ \nu^*_-(t)\nu_-(t)\right),
\end{eqnarray}
where we have defined
\begin{eqnarray}
\nu_+(t) &=& 2e^{-iE_5t}  - e^{-iE_{9}t} - e^{-iE_{13}t}, \nonumber \\
\nu_-(t) &=& 2e^{-iE_7t}  - e^{-iE_{11}t} - e^{-iE_{15}t}.
\end{eqnarray}
From the eigenenergies (\ref{eq:ener}) it becomes clear that $p_{\downarrow \downarrow}
=p_{\uparrow \uparrow}$.
Figure (\ref{fig:evo3}) depicts the calculated probabilities.
We again get time-scaling with the coupling strength.

\subsection{Bell state $|\Psi^-\rangle$}
This last case corresponds to $B_-=1, A_+=A_-=B_+=0$ in (\ref{eq:genstate}); it actually
is the simplest one because the initial four qubit state equals a basis state.
Explicity we obtain
\begin{eqnarray}
|\psi_-\rangle &=& - |e_2\rangle,  \nonumber \\
|\psi_-(t) \rangle &=& - e^{-iE_2t} |e_2\rangle,  \nonumber \\
P_{\uparrow \uparrow}|\psi_-(t) \rangle &=& - e^{-iE_2t} |S^{[12]}\rangle \otimes |T_+^{[ab]}\rangle 
\end{eqnarray}
and the concomitant trivial probabilities
\begin{equation}
p_{\uparrow \uparrow}=1, \quad p_{\uparrow \downarrow}= p_{\downarrow \uparrow}= p_{\downarrow \downarrow}=0.
\end{equation}
After factoring out the ancillas state, the two-qubit state is -up to a phase- equal to the target
state. As a consequence, the fidelity is equal to one. The unitary evolution has no negative impact
on the ideal probabilities and fidelity for this particular Bell state.

\section{Quantum gates by unitary evolution}\label{sec:Gate}
The obtained results show that the probabilities which were initially one, first decrease but
after a time $\tilde t$ again become equal to one. For the Bell state $|\Phi^-\rangle$, we note
that for $t=\tilde t/2$ the up-down probability (initially one) becomes zero, whereas the
down-up probability equals one. This opens the possibility for two-qubit gates, that is an
operation on
the data qubits, which can be verified by measuring the ancillas without affecting the
state of the data qubits. The time $\tilde t$ follows from 
\begin{equation}
\mathcal J \tilde t = 2 \pi \quad \text{as} \quad \tilde t= \frac{2\pi}{\mathcal J}.
\end{equation}
In order to address two-qubit gates in the four-qubit system
we continue the analysis started above.
To this end, we also need the computational basis. For two qubits
we define the basis states $|c_k\rangle, k=1,2,3,4$ as the common computational states
\begin{equation}
|c_1\rangle =\ketdd, |c_2\rangle =\ketdu, |c_3\rangle =\ketud, |c_4\rangle =\ketuu.
\end{equation}
The computational states for four qubits are the product states
\begin{equation}
|c_{kl}\rangle =|c_k^{[12]}\rangle \otimes |c_l^{[ab]}\rangle, \quad k,l=1,2,3,4.
\end{equation}
We relabel the ancilla measurement operators (\ref{eq:meop}) also as $P_l$ and rewrite
these as
\begin{equation}
P_l^{[ab]} = \mathcal I^{12} \otimes |c_{l}^{[ab]}\rangle \langle c_l^{[ab]} |
= \sum_{j=1}^4 |c_{jl}\rangle \langle c_{jl}|.
\end{equation}
The initial state is assumed to be the product state of a computational basis
state of the ancillas, say $|c_n^{[ab]}\rangle$ with fixed $n$  and a general two-data qubit state:
\begin{equation}
|\psi_0 \rangle= \left\{\sum_{m=1}^4 \alpha_m  |c_{m}^{[12]} \rangle \right\} \otimes |c_n^{ab}\rangle =
 \sum_{m=1}^4 \alpha_m  |c_{mn} \rangle.
\label{eq:instate}
\end{equation}
After unitary evolution governed by the operator (\ref{eq:evo}),
the state is given by
\begin{equation}
|\psi(t) \rangle= \sum_{k=1}^{16} \sum_{m=1}^4 \alpha_m \langle e_k |c_{mn} \rangle
e^{-iE_kt} |e_k \rangle.
\end{equation}
Applying the ancilla measurement operator $P^{[ab]}_l$ to this state yields
\begin{eqnarray}
P^{[ab]}_l |\psi(t) \rangle &=& \sum_{j=1}^{4} \sum_{m=1}^4 \alpha_m 
W_{jl-mn}(t) |c_{jl} \rangle \\
&=& \left\{ \sum_{j=1}^{4} \sum_{m=1}^4 \alpha_m 
W_{jl-mn}(t) |c_{j}^{[12]} \rangle \right\} \otimes |c_l^{[ab}\rangle,\nonumber
\end{eqnarray}
where we have defined
\begin{equation}
W_{jl-mn}(t) =
\sum_{k=1}^{16}  \langle e_k |c_{mn} \rangle
 \langle c_{jl}  |e_k \rangle e^{-iE_kt}.
\end{equation}
Since the coefficients 
$\langle e_k |c_{mn} \rangle$ are real, $W$ is symmetric: 
$W_{jl-mn}(t) = W_{mn-jl}(t)$.
The measurement probabilities for obtaining the result $l$ follow as
\begin{equation}
p_{ln}(t) = 
\sum_{m=1}^{4} \sum_{m'=1}^{4} \sum_{j=1}^{4} 
\alpha^*_{m'} \alpha_m 
W^*_{jl-m'n}(t) W_{jl-mn}(t) ,
\label{eq:pln}
\end{equation}
where we have explicitly indicated the dependence on the initial ancillas state.
Note the implicit dependence on the initial data qubit states.
The corresponding amputated and normalized final data qubit state is obtained as
\begin{eqnarray}
|\psi_{ln}^{[12]}(t) \rangle &=& \frac{1}{\sqrt{p_{ln}}}
\sum_{j=1}^{4} \sum_{m=1}^4 \alpha_m 
W_{jl-mn}(t) |c_{j}^{[12]} \rangle \nonumber\\ &=&
\sum_{j=1}^4 
\beta^{[ln]}_j(t) |c_{j}^{[12]} \rangle , 
\end{eqnarray}
with the functions
\begin{equation}
\beta^{[ln]}_j(t) = \frac{1}{\sqrt{p_{ln}}} \sum_{m=1}^{4} \alpha_m W_{jl-mn}(t).
\end{equation}
Comparison of this final state and the initial one (\ref{eq:instate}) suggest a linear relation between
these states. Because of the abovementioned dependence of the probabilities on the
coefficients $\alpha_m$ this is, however, in general not true.
The dependence only vanishes for zero probability which is not interesting
and for the relevant case of probability one.
If for an instant of time $\tilde t$ and a specific combination of $l, n$
\begin{equation}
\sum_{j=1}^{4} W^*_{jl-m'n}(\tilde t) W_{jl-mn}(\tilde t) = \delta_{mm'},
\label{eq:constr}
\end{equation}
the probability becomes one
\begin{equation}
p_{ln}(\tilde t) = \sum_{m=1}^{4} \sum_{m'=1}^{4}
\alpha^*_{m'} \alpha_m \delta_{mm'}
 = \sum_{m=1}^{4} \alpha^*_m \alpha_m =1.
\end{equation}
In fact, in the numerical implementation described below, we have found that the specific $l,n$ combinations
are $l=n$. Thus the latter two equations actually read
\begin{equation}
\sum_{j=1}^{4} W^*_{jl-m'n}(\tilde t) W_{jl-mn}(\tilde t) = \delta_{mm'} \delta_{ln},
\label{eq:constr-ln}
\end{equation}
and
\begin{equation}
p_{ln}(\tilde t) = \delta_{ln} \sum_{m=1}^{4} \sum_{m'=1}^{4}
\alpha^*_{m'} \alpha_m \delta_{mm'}
 =  \delta_{ln}\sum_{m=1}^{4} \alpha^*_m \alpha_m = \delta_{ln}.
\end{equation}
It indeed implies that the coefficients
$\beta_j^{[nn]}(\tilde t)$ depend linearly on the initial ones $\alpha_m$.
Consequently, we obtain a final two-qubit state which follows as a gate operation
on the initial one
\begin{eqnarray}
|\psi_{nn}^{[12]}(\tilde t) \rangle &=& \sum_{j=1}^4
\beta^{[nn]}_j(\tilde t) |c_{j}^{[12]} \rangle \nonumber \\ &=&
\sum_{j=1}^{4} \sum_{m=1}^4 \alpha_m
W_{jn-mn}(\tilde t) |c_{j}^{[12]} \rangle.
\end{eqnarray}
In matrix form, the gate is explicitly given by
\begin{equation}
\beta^{[nn]}_j(\tilde t) =  \sum_{m=1}^4 
%G_{jm}^{[ln]}(\tilde t)
 W_{jn-mn}(\tilde t)  \alpha_m.
\end{equation}
%with matrix
%\begin{equation}
 %G_{jm}^{[ln]}(\tilde t) = W_{jl-mn}(\tilde t).
%\end{equation}
Unitarity of the gate, or matrix, is guaranteed by the constraints
(\ref{eq:constr},\ref{eq:constr-ln}).
%and also requires  
%\begin{equation}
%\sum_{m=1}^{4} W^*_{j'l-mn}(\tilde t) W_{jl-mn}(\tilde t) = \delta_{jj'}.
%\label{eq:constr2}
%\end{equation}
%It is actually not {\it a priori} clear that the constraints (\ref{eq:constr})
%and (\ref{eq:constr2}) are independent.
%Here, it is also  numerically found that only combinations with $l=n$ do the job.
%Therefore we augment this 
%relation as
%\begin{equation}
%\sum_{m=1}^{4} W^*_{j'l-mn}(\tilde t) W_{jl-mn}(\tilde t) = \delta_{jj'} \delta_{nl}.
%\end{equation}
%Though we have not proved this,
%we suspect the appearance of the second Kronecker delta to be a consequence of
%the ancilla-- data qubit symmetry and unitarity in the
%four qubit Hilbert space.
Note that, for an arbitrary initial data qubit state,
the probability only becomes one 
for obtaining the {\it same} ancilla spin values as initialized.
In situations where the data qubits are initialized in a specific
computational state, other measurement results, e.g. $\downarrow  \uparrow$
instead of $\uparrow \downarrow$,  with probability one also have appeared.
It does not simultaneously happen, however, for all four initial computational data qubit states and,
consequently, no unitary operation is effectuated.

In the way described above, four unitary gates are in principle obtained by evolution and projection
at $t = \tilde t$:
\begin{equation}
 U_{jm}^n(\tilde t) = W_{jn-mn}(\tilde t), \quad n=1,2,3,4.
\end{equation}
It will further be analyzed numerically as well as analytically; these methods
have turned out to be complementary for obtaining the results.

\subsection{Implementation}
\begin{figure}[htb] 
\centering
\includegraphics[width=7.9cm]{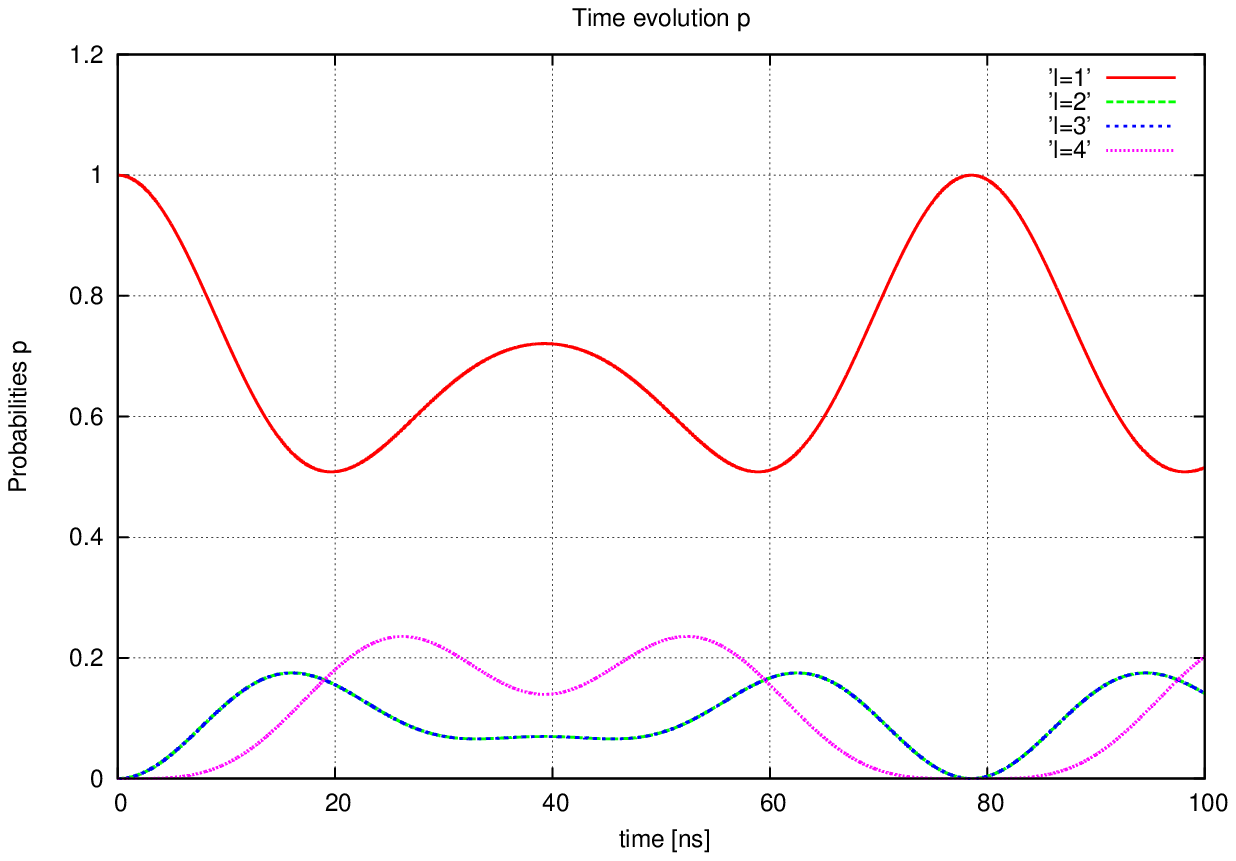}
\includegraphics[width=7.9cm]{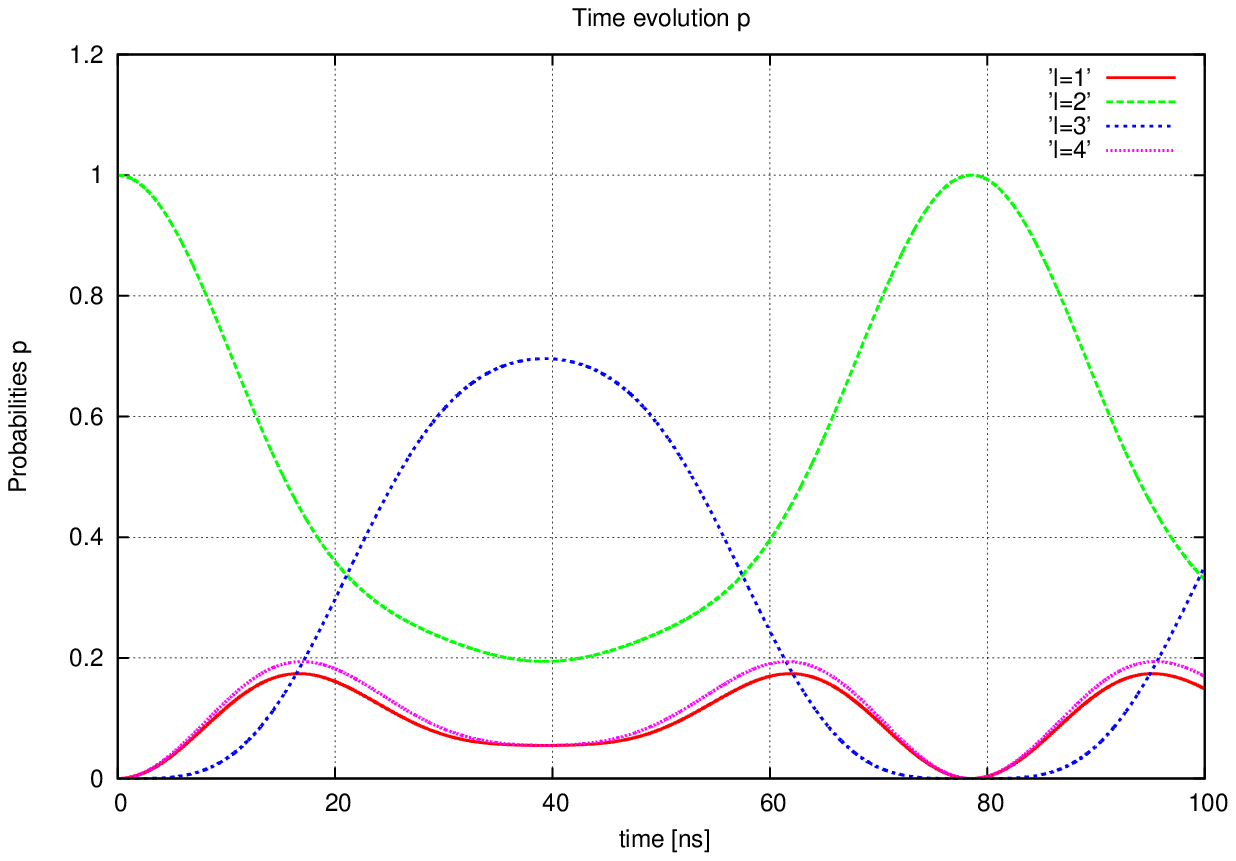}
\caption{\label{fig:gate} Probabilities: $\mathcal J=$ 80 MHz; top $n=1$, low $n=2$ }
\end{figure}

We have implemented the formalism in a computer program. To compute the
quantities $W_{jl-mn}(t)$, all the necessary ingredients have been obtained analytically.
These are the eigenstates $|e_k \rangle$ and the concomitant energy eigenvalues $E_k$ presented in section
\ref{sec:four}.
Additionally, the basis transformation $\langle e_k| c_{mn} \rangle$ is needed.
Since the eigenstates are expressed in singlet-triplet basis states -for data and ancilla qubits-
it can be computed using the transformation between singlet-triplet and computational basis.
%The explicit computation is relegated to  appendix \ref{app:basis}.
Next the measurement probabilities (\ref{eq:pln}) are  calculated for given
initial data qubit state, {\it i.e.}, given coefficients $\alpha_m$. They can be chosen by hand
or randomly generated by a random-number generator for a given probability density function. We have chosen
a uniform distribution around zero for real as well as imaginary parts. Of course, the state
has then to be normalized. The resulting measurement probabilities for all $l,n$
combinations are plotted as a function of time and the instant of time where a probability becomes
one is identified. Indeed we find that for $l=n$ this instant of time is given by
$\tilde t= \frac{2\pi}{\mathcal J}$.
Figure (\ref{fig:gate}) shows the measurement probabilities for $n=1$ and $n=2$ and all $l=1,2,3,4$.
It explicitly illustrates the already mentioned results. The initial data qubit state
 is randomly selected in this example.

The resulting gates  explicitly read
\begin{eqnarray}
\label{eq:gates}
U^1&=&
\begin{pmatrix}
e^{\frac{4\pi i\omega}{\mathcal J}} & 0 & 0 & 0\\
0 & e^{\frac{2\pi i\omega}{\mathcal J}} & 0 & 0 \\
0 & 0&  e^{\frac{2\pi i\omega}{\mathcal J}} & 0  \\
0 & 0 & 0 & 1 
\end{pmatrix}, \nonumber \\
U^2 &=&  e^{-\frac{2\pi i\omega}{\mathcal J}} U^1, \quad
U^3 =  e^{-\frac{2\pi i\omega}{\mathcal J}} U^2, \nonumber \\
  U^4&=& e^{-\frac{2\pi i\omega}{\mathcal J}} U^3.  
\end{eqnarray}
These are four, phase-equivalent phase gates.
%Note that the expressions are
%computed numerically - we just put in the analytical expression for convenience.

\subsection{Analytical construction}
With some effort, these gates can be obtained analytically as well.  
The inititial state is completely specified as a product state of
a computational data qubit state and a computational ancilla qubit state.
It is rewritten in terms of eigenstates of the Hamiltonian in order to
construct the time-evoluted state.  Next the time is identified upon a measurement
yields a result with probability one and the resulting two-data-qubit state is identified.
Let is explicitly demonstrate this for the initial state
\begin{equation}
|\phi_0\rangle = \ketdd \otimes \ketdu,
\end{equation}
which reads in terms of singlet-triplet basis states as
\begin{equation}
|\phi_0\rangle = \tfrac{1}{2}\sqrt{2} |T_-^{[12]}\rangle\left(|T_0^{[ab]}\rangle-|S^{[ab]}\rangle \right).
\end{equation}
Rewriting it as a linear combination of eigenstates yields
\begin{equation}
|\phi_0\rangle = -\tfrac{1}{2}\left(|e_{11}\rangle-|e_{15}\rangle
+\sqrt{2} |e_7\rangle \right).
\end{equation}
As a consequence, after time evolution we get
\begin{equation}
|\phi(t)\rangle = -\tfrac{1}{2}\left(e^{-iE_{11}t} |e_{11}\rangle-e^{-iE_{15}t}|e_{15}\rangle
+\sqrt{2} e^{iE_7t} |e_7\rangle \right).
\end{equation}
Now it is readily checked that
\begin{eqnarray}
|\phi(\tilde t)\rangle &=& -\tfrac{1}{2}\left(e^{i\omega\tilde t} |e_{11}\rangle-e^{i\omega\tilde t}|e_{15}\rangle
+\sqrt{2} e^{i \omega \tilde t} |e_7\rangle \right) \nonumber \\ &=& e^{i\omega\tilde t} |\phi_0\rangle,
\end{eqnarray}
which implies $p_{\downarrow,\uparrow}=1$ for $t=\tilde t$. Thus the resulting change
in the data two qubit state can be identified as  
\begin{equation}
\ketdd \rightarrow e^{\frac{2\pi i \omega}{\mathcal J}} \ketdd = U^2_{11} \ketdd.
\end{equation}
At this point it also follows that $U^2_{1m}=0$ for $m=2,3,4$. In this way,
all four gates given in (\ref{eq:gates}) can be constructed.

\subsection{Universality}
Naturally the question arises whether the $U^k$ are universal gates. Since they are phase-equivalent, 
we just consider $U^2$. The unitary matrix $U^2$ acts non-trivially on two vector components because
it is diagonal with two elements equal to one. 
Consequently, it is a two-level unitary matrix \cite{NC}.
Any unitary matrix can be decomposed in terms of two-level unitary matrices. In \cite{NC},
this fact is used to prove universality of CNOT-gates and single qubit gates. The reasoning is
that with these gates one may construct a arbitrary two-level unitary operation
and these, in turn, can be used to build any unitary operation.
However, generally a number of two-level unitary matrices is necessary.
Thus the fact that $U^2$ is two-level unitary operation does {\it not}
imply that $U^2$ (and single qubit gates) is indeed sufficient universal for quantum computing.

It is actually easy to see that $U^2$ can be written as a product of single qubit rotations:
\begin{equation}
R^{[1]}_z(\beta) \otimes R^{[2]}_z(\beta)  =
\begin{pmatrix}
e^{- i \beta} & 0 & 0 & 0\\
0 & 1 & 0 & 0\\
0 & 0 & 1 & 0\\
0 & 0 & 0 & e^{ i \beta} 
\end{pmatrix},
\end{equation}
which is equal to $U^2$ for $\beta= -2\pi i \omega/\mathcal J$.

Nevertheless we believe it to be elucidating to
demonstrate that $U^2$ can be written in terms of CNOT-like operations and
a controlled single qubit rotation (as is done in \cite{NC} for a different example).
Instead of the standard CNOT we need the controlled operation with the NOT gate being
performed on the second qubit, conditional on the first qubit being in the ground state, cf. \cite{NC} Figure (4.11).
Denoting this gate $C^0$, we explicitly have in the computational basis
\begin{equation}
C^0 = 
\begin{pmatrix}
0 & 1 & 0 & 0 \\
1 & 0 & 0 & 0 \\
0 & 0 & 1 & 0 \\
0 & 0 & 0 & 1 
\end{pmatrix}.
\end{equation}
Next we use the  single qubit operation $\tilde U$ which is applied to the first qubit conditional
on the second qubit (being excited); the 2-qubit operation is denoted by $C_{\tilde U}$. These matrices
read
\begin{equation}
\tilde U  = 
\begin{pmatrix}
e^{\frac{2\pi i \omega}{\mathcal J}} &  0 \\
0& e^{-\frac{2\pi i \omega}{\mathcal J}} 
\end{pmatrix}, \quad
C_{\tilde U} = 
\begin{pmatrix}
1 & 0 & 0 & 0 \\
0 & e^{\frac{2\pi i \omega}{\mathcal J}} & 0 & 0  \\
0 & 0 & 1 & 0 \\
0 & 0 & 0 &  e^{-\frac{2\pi i \omega}{\mathcal J}}
\end{pmatrix}.
\end{equation}
By matrix multiplication it can easily be verified  that
\begin{equation}
U^2 = C^0 \, C_{\tilde U}  \, C^0,
\end{equation}
thereby having achieved the desired decomposition. Note that $C_{\tilde U}$ is
a controlled rotation around $z$-axis.

\subsection{Discussion}
The ancilla measurement at $t=\tilde t$ should yield with probability one the same spin values
for the ancillas
as in the initialization. Therefore it serves as a check whether the gate operation on data qubits
has been performed correctly without affecting this data qubit state. A different ancilla measurement
result indicates an error. In this way `verifiable' phase gates are effectuated.
In practice, an error can nevertheless not be excluded.

In an actual application, it is of course not necessary to measure the ancillas at $t=\tilde t$.
At that moment, the complete four-qubit state is a product state of the state of the ancillas
and a two data-qubit state which is predictable. Moreover, its evolution starting from
an arbitrary state is described by a unitary phase gate.
Switching off the interaction $\mathcal J$ at $t=\tilde t$
ensures that data qubits and ancillas then remain decoupled and the subsequent dynamics is only
governed by free evolution. 

Above, we have already addressed the topic of universality of the constructed $U^k$ gates. 
For universal quantum computing, we would need to express the CNOT-gate
or the CZ gate in terms of the considered gate and single qubit operations.
Since the $U^2$ phase gate can be written as the product of two single qubit rotations,
this is not possible.
%We have, however, serious doubts whether this is possible  but cannot exclude
%it at this point.
Nevertheless, the found gates  may be applicable in quantum computation 
if complemented by additional operations. The controlled $Z$-rotation $C_{\tilde U}$ can actually
be useful in the quantum Fourier transformation.

Suppose one measures the ancillas at another moment where there are generally four non-zero
probabilities for the possible outcomes. The resulting state is again a product state of 
the two data qubit-state and the computational two ancilla state corresponding to the measurement
result. In other words, data-qubits and ancillas are disentangled and therefore 
a two-data qubit state can once more be identified. Its coefficients, however, do
not depend linearly on the initial amplitudes. Therefore, the change cannot be interpreted
as a unitary data-qubit gate. This is due to the projective measurement
of the ancillas. As we have seen, such an interpretation is valid for a measurement
with probability one.
%The data-qubits are dis-entangled from the ancillas at that instant of time.

Another observation concerning the gate construction
is the necessity of the ancillas being initialized in one of the
computational states. If the initial state of the ancillas is a superposition, we
cannot identify an instant of time where a specific measurement result is certain.
Once more, this prohibits a unitary two-data qubit gate interpretation.

%Finally, we mention that we have also initiated a study of analogous $XX$  and $ZX$
%measurements and concomitant initialization of the ancillas. The aim is to possibly
%find more two-qubit gates.  The preliminary results often show
%very fast oscillations in measurement probabilities. This is due to 
%$H_0$, that is the coupling of the strong magnetic field to the spins, yielding
%high frequent Larmor precession for $S_x$ and $S_y$.
%As is argued in \cite{Naus-spin} these high frequency effects  do not
%vanish in a rotating frame in such cases. It prohibits the implementation
%of a possibly `hidden' (slower) gate. Note, however, that we have not finished
%this study yet and these remarks are preliminary.

\section{Conclusion}
Four electron spin qubits in quantum dots subject to a magnetic field have been modelled by a standard
exchange Hamiltonian. Two data qubits and two ancilla qubits are considered; the exchange
interaction is only present between a data-qubit and an ancilla. The time-independent symmetric case
is solved by constructing the eigenbasis of the Hamiltonian and computing the corresponding energy eigenvalues.

As an application,
the resulting evolution operator is exploited to obtain the time evolution of selected states.
These states are product states of data-qubit Bell states and specific ancilla basis states. They are
known to be the stable output states of the $X$ and $Z$ stabilizing circuit  \cite{Fowler}.
Unitary evolution of the stable states before
the measurement of the ancillas is calculated.
Resulting probabilities for
the error syndrome and fidelities can be obtained. They are seen to differ substantially from
the non-evolving case. As extensively discussed in \cite{Naus-stab}, it may lead to reconsideration
of the proposed error correction and surface codes for fault-tolerant computation.

The present study, however, indicates that
this possible problem is less serious for electron spin qubits than for
superconducting qubits coupled to resonators considered in \cite{Naus-stab}.
We have shown for the considered model that Hamiltonian unitary 
evolution scales with the exchange coupling. Hence, it suggests to tune down the
exchange interaction in the idle periods of time, {\it i.e.}, if it is not required.
For coupling $\mathcal J = 800$ MHz, we have obtained a typical time interval of 4 ns
for probabilities to change from zero to one and vice versa. Decreasing the coupling to
$\mathcal J \approx $ 200 kHz, a realistic minimum value \cite{Syp}, yields for that time
interval $16 \mu$s, which excludes unwanted rapid changes in probabilities and fidelities.

The possibility of verifiable two-qubit phase gates due to unitary evolution is shown.
The verification is done
by means of ancilla measurement. After the prescribed elapsed evolution time,
the probabilities for finding the respective ancilla measurement results
are one and zero. Consequently, obtaining the correct measurement result
confirms the realization of the two-qubit operation without disturbing the final data-qubit state.
An explicit decomposition of the found phase gate in terms of a CNOT and controlled single qubit
rotations is given. It can also be written as  product of two single qubit rotations
and is therefore no alternative for controlled two-qubit operations in quantum computing.

Nevertheless, the verifiable two-qubit gate may have some benefits as
its verifiability. Furthermore, the unitary gate governed by Hamiltonian evolution is equivalent
to {\it two} single qubit operations induced by microwave pulses. 
These possible merits have to be assessed in future experiments
with electron spin devices.

Large scale fault-tolerant quantum computation is the long-term goal of
ongoing research for developing quantum computers using various qubits. The 
considered four electron spin qubit system may be used as building block, or unit cell,
in such an extended electron spin quantum device. Of course, these unit cells need to be
eventually coupled by, for example, electromagnetic resonators.
The symmetry of the unit cells, characterized by their total spin, appears to be
advantageous for analyzing such possible developments.

\section*{Acknowledgments}
The author would like to thank G.J.N. Alberts, A.-J. de Jong, T. Last, L.M.K. Vandersypen and
R. Versluis for stimulating discussions and/or
a critical reading of the manuscript.
This research is supported by the Early Research Programme of the Netherlands
Organisation for Applied Scientific Research (TNO).
Additional support from the
Top Sector High Tech Systems and Materials is highly appreciated.


\begin{thebibliography}{20}

\bibitem{Engel} H.-A. Engel, L.P. Kouwenhoven, D. Loss, and C.M. Marcus,
{\em Spins in few-electron quantum dots}, Quant. Infor. Proc. {\bf 3}, 115 (2004).

\bibitem{Hanson} R. Hanson, L.P. Kouwenhoven, J.R. Petta, S. Tarucha and L.M.K.
Vandersypen, {\em Spins in few-electron quantum dots}, Rev. Mod. Phys. {\bf 79}, 1217 (2007).

\bibitem{Kawa} E. Kawamaki, {\em Characterization of an electron spin qubit in a Si/SiGe quantum dot},
Ph. D. thesis, TUDelft (2016).

\bibitem{Loss98} D. Loss and D.P. DiVincenzo, {\em Quantum computation with
quantum dots}, Phys. Rev. A  {\bf 57}, 120 (1998).

\bibitem{Loss00} D. Loss, G. Burkard and D.P. DiVincenzo, {\em Electron spins
in quantum dots as quantum bits}, Journal of Nanoparticle Research  {\bf 2}, 401 (2000).

\bibitem{Ash} N.W. Ashcroft and N.D. Mermin, {\em Solid State Physics}, International Edition,
Saunders College HRW (1976).

\bibitem{Meun} T. Meunier, V.E. Calado and L.M.K. Vandersypen, {\em Efficient
controlled-phase gate for single-spin qubits in quantum dots}, Phys. Rev. B
{\bf 83}, 121403(R) (2011).

\bibitem{Watson} T.F. Watson {\it et al.}, {\em A programmable two-qubit quantum processor in silicon},
Letter Nature {\bf 555}, 633 (2018).

%\bibitem{Merz} E. Merzbacher, {\em Quantum Mechanics}, Second Edition, John Wiley \& Sons, Inc.,
  %New York, (1970).
%
  %\bibitem{Cohen1} C. Cohen-Tannoudji, B. Diu, F. Lalo\"e, {\em Quantum Mechanics},
    %Volume {\bf 1}, Hermann, Paris, and Wiley-VCH (2005).

  \bibitem{Fowler} A.G. Fowler, M. Mariantoni, J.M. Martinis, and A.N. Cleland, {\em Surface
codes: Towards practical large-scale quantum computation}, Phys. Rev. A {\bf 86},
032324 (2012).

\bibitem{Naus-stab}
H.W.L. Naus and R. Versluis, {\em
Consequences of unitary evolution of coupled qubit-resonator systems
for  stabilizing circuits in surface codes}
arXiv:{\bf 1811.09832} [quant-ph] (2018).

  \bibitem{Cohen2} C. Cohen-Tannoudji, B. Diu, F. Lalo\"e, {\em Quantum Mechanics},
    Volume {\bf 2},  Hermann, Paris, and Wiley-VCH (2005).

\bibitem{NC} M.A. Nielsen and I.L. Chuang, {\em Quantum Computation and
  Quantum Information}, Cambridge University Press, Cambridge, UK  (2000).

\bibitem{Syp}  L.M.K. Vandersypen, {\it private communication}.
%\bibitem{Kopp} F.H.L. Koppens, {\em Coherence and control of a single electron spin qubit in a quantum dot},
%Thesis, TUDelft (2007).
%
%\bibitem{Schoel} R.J. Schoelkopf and S.M. Girvin, {\em Wiring up quantum systems},
%%Nature {\bf 451}, 664 (2008).

%\bibitem{Taka} M. Takahashi, S.D. Bartlett and A.C. Doherty, {\em Tomography
%of a spin qubit in a double quantum dot}, Phys. Rev. A  {\bf 88}, 022120 (2013).
%
%\bibitem{Levi} L.S. Levitov and E.I. Rashba, {\em Dynamical
%spin-electric coupling in a quantum dot}, Phys. Rev. B  {\bf 67}, 115324 (2003).
%
%\bibitem{Toku} Y. Tokura, W.G. van der Wiel, T. Obata and S. Tarucha, {\em Coherent
%single electron spin control in a slanting Zeeman field }, Phys. Rev. Lett.  {\bf 96}, 047202 (2005).
%
%\bibitem{Elzer} J.M. Elzerman, R. Hanson, L.H. Willems van Beveren, B. Witkamp,
%L.M.K. Vandersypen and L.P. Kouwenhoven {\em Single-shot read-out of an
%individual electron spin in a quantum dot},
%Nature {\bf 430}, 431 (2004).
%
%\bibitem{Petta} J.R. Petta {\it et al.}, {\em Coherent manipulation
%of coupled electron spins in semiconductor quantum dots},
%Science {\bf 309}, 2180 (2005).
%
%\bibitem{Zajac15} D.M. Zajac, T.M. Hazard, X. Mi, K. Wang
%and J.R. Petta, {\em A reconfigurable gate architecture for Si/SiGe quantum dots},
%Applied Phys. Lett. {\bf 106}, 223506 (2015).
%
%\bibitem{Zajac16} D.M. Zajac, T.M. Hazard, X. Mi, E. Nielsen
%and J.R. Petta, {\em Scalable gate architecture for densely packed semiconductor spin qubits},
%Phys. Rev. Applied {\bf 6}, 054013 (2016).
%
%\bibitem{Zajac17} D.M. Zajac, A.J. Sigillito, M. Russ, F. Borjans, J.M. Taylor, G. Burkard
%and J.R. Petta, {\em Quantum CNOT gate for spins ins silicon},
%arXiv: {\bf 1708.03530} [cond-mat.mes-hall] (2017).
%
%\bibitem{Stehlik} J. Stehlik {\it at al.}, {\em Fast charge sensing of a cavity-coupled double quantum dot
%using  a Josephson parametric amplifier}, Phys. Rev. Appl. 4, 014018 (2015).
%
%\bibitem{Veld14} M. Veldhorst {\it et al.},
%{\em An addressable quantum dot qubit with fault-tolerant control fidelity},
%Nature Nanotechnology {\bf 9}, 981 (2014).
%
%\bibitem{Schmidt} A.R. Schmidt {\it et al.},
%{\em A prototype silicon double quantum dot qubit with dispersive microwave readout},
%Journ. of Applied Phys. {\bf 116}, 044503  (2014).
%
%\bibitem{Sar} S. Das Sarma, X. Wang and S. Yang,
%{\em Hubbard model description of silicon spin qubits: charge stability diagram and
%tunnel coupling in Si quantum dots}, Phys. Rev. B {\bf 83}, 235314 (2011).
%
%\bibitem{Yang} S. Yang, X. Wang and S. Das Sarma,
%{\em Generic  Hubbard model description of semiconductor quantum dot spin qubits},
%Phys. Rev. B {\bf 83}, 161301 (2011).
%
%\bibitem{Hens} T. Hensgens, {\em Quantum simulation of a Fermi-Hubbard model using a 
%semiconductor quantum dot array}, Nature {\bf 548}, 70 (2017).
%
%\bibitem{Veld16} M. Veldhorst, H.G.J. Eenink, C.H. Yang and A.S. Dzurak,
%{\em Silicon CMOS architecture for spin-based quantum computer},
%arXiv: {\bf 1609.09700} [cond-mat.mes-hall] (2016).
%
%\bibitem{Wei13} H.-R. Wei and F.-G. Deng, {\em Universal quantum gates on electron-spin qubits
%with quantum dots inside single-side optical microcavities}, Optics Express {\bf 22}, 593 (2013).
%
%\bibitem{Wei14} H.-R. Wei and F.-G. Deng, {\em Scalable quantum computing based on
%stationary spin qubits in coupled
%quantum dots inside double-sided optical microcavities}, Scientific Reports {\bf 4: 7551} (2014).
%
% \bibitem{AS} M. Abramowitz and I.A. Stegun, {\em Handbook of mathematical functions}, 10th printing,
%%   Dover, New York (1972).
%%
%% \bibitem{GR} I.S. Gradshteyn and I.M. Ryzhik, {\em  Table of Integrals, Series, and Products},
%%   Corrected and enlarged edition, Academic Press, New York, (1980).
%%
%% \bibitem{Kampen} N.G. van Kampen, {\em Stochastic Processes in Physics and Chemistry},
%%   Third edition, North-Holland Personal Library, Elsevier, Amsterdam (2007).
%
\end{thebibliography}
\end{document}